\documentclass[aps,prl,notitlepage,superscriptaddress,showpacs,twocolumn ]{revtex4-1}
\usepackage{graphicx,subfigure,epsfig}
 \usepackage{array,multirow}
\usepackage{dcolumn}
\usepackage{amssymb,amsmath,amsfonts,mathrsfs}
\usepackage{array}
\usepackage{times,setspace}
\usepackage{latexsym}
\usepackage{float,flafter,bm,bbm}
\usepackage{epstopdf,color,multirow}
\usepackage[colorlinks,linkcolor=blue,anchorcolor=blue,urlcolor=blue,citecolor=blue]{hyperref}
\usepackage{footnote}

\hypersetup{
    colorlinks=true,
    linkcolor=blue,
    filecolor=magenta,
    urlcolor=blue,
}

\begin{document}
\title{Phase Diagram of the Square-Lattice $t$-$J$-$V$ Model for Electron-Doped Cuprates}

\author{Qianqian Chen}
\affiliation{Kavli Institute for Theoretical Sciences, University of Chinese Academy of Sciences, Beijing 100190, China}
\author{Lei Qiao}
\affiliation{Kavli Institute for Theoretical Sciences, University of Chinese Academy of Sciences, Beijing 100190, China}

\author{Fuchun Zhang}
\email{fuchun@ucas.ac.cn}
\affiliation{Kavli Institute for Theoretical Sciences, University of Chinese Academy of Sciences, Beijing 100190, China}
\affiliation{CAS Center for Excellence in Topological Quantum Computation, University of Chinese Academy of Sciences, Beijing, 100190, China}

\author{Zheng Zhu}
\email{zhuzheng@ucas.ac.cn}
\affiliation{Kavli Institute for Theoretical Sciences, University of Chinese Academy of Sciences, Beijing 100190, China}
 \affiliation{CAS Center for Excellence in Topological Quantum Computation, University of Chinese Academy of Sciences, Beijing, 100190, China}

\begin{abstract}
Motivated by significant discrepancies between experimental observations of electron-doped cuprates and numerical results of the Hubbard and $t$-$J$ models,
we investigate the role of inter-site interactions $V$ by studying the $t$-$J$-$V$ model on square lattices. 
Based on large-scale density matrix renormalization group simulations, we identify the ground-state phase diagram across varying inter-site interactions $V$ and doping concentration $\delta$. We find that the phase diagram with finite inter-site {interactions} $2\lesssim V/J\lesssim3$ offers a more accurate description of electron-doped cuprates than the conventional Hubbard and $t$-$J$ models. 
Moreover, we reveal the role of inter-site interactions $V$ at varying doping levels: at light doping, inter-site interactions favor N\'{e}el antiferromagnetic order, and suppress both superconductivity and charge density wave; 
{around optimal} doping, these interactions support a pseudogap-like phase while suppressing superconductivity, and we further perform the slave boson mean-field analysis to understand the numerical results microscopically; at higher doping, the effects of inter-site interactions become insignificant, with our numerical predictions suggesting the emergence of incommensurate {spin density wave} phase.  
Our specific focus around optimal doping with various inter-site interactions  
identifies successive phases including phase separation, 
uniform $d$-wave SC and a pseudogap-like phase, and reveals a relative insensitivity of charge density wave to superconductivity. 
Our study suggests the $t$-$J$-$V$ model as the minimal model to capture the essential physics of the electron-doped cuprates. 

\end{abstract}

\maketitle

\emph{Introduction.---}
Understanding the intricate physics of high-temperature superconductors copper-oxide stands as one of the great challenges in modern condensed matter physics \cite{Bednorz1986, Imada1998,Orenstein2000,PatrickLee2006, Armitage2010,Keimer2015}. Theoretically, this foundational puzzle was cast within the framework of the Hubbard or $t$-$J$ model \cite{FCZhang1988,Anderson2004,PatrickLee2006, Keimer2015, Arovas2022, MingpuQin2022}. 
Nonetheless, there remain persistent discrepancies within the parameter range most applicable to cuprates, 
stemming from the mismatch between experimental observations and state-of-the-art numerical simulations based on the Hubbard or $t$-$J$  models~\cite{FuchunZhang1989,Zhu2014Nature,ZhengZhu2018, ShuaiChen2018, Dodaro2017, HongChenJiang2019, YiFanJiang2020, MingpuQin2020, ChiaMinChung2020, ShoushuGong2021, ShengtaoJiang2021, ShengtaoJiang2022, XinLu2023, HaoXu2023, XinLu2023Sign}.
Specifically, on the hole-doped side, experimental findings reveal a notable superconducting  feature~\cite{Lee06,Scalapino2012}, whereas the existence of robust superconductivity (SC) remains elusive in numerical simulations~\cite{Dodaro2017,HongChenJiang2019,YiFanJiang2020, MingpuQin2020, ChiaMinChung2020,ShengtaoJiang2021, ShengtaoJiang2022, XinLu2023, HaoXu2023, XinLu2023Sign, FengChen2023}. {Furthermore, simulating the original three-band model for hole-doped cuprates poses additional computational challenges due to its complexity.
Conversely, in the electron-doped case, noteworthy discrepancies regarding SC and N\'{e}el antiferromagnetic order (AF) exist between the findings of   experiments~\cite{Takagi1989, Uefuji2001, Motoyama2007, Armitage2010, Scalapino2012} and unbiased numerical studies based on the Hubbard or $t$-$J$  models \cite{ShoushuGong2021, ShengtaoJiang2021, ShengtaoJiang2022, XinLu2023, HaoXu2023, FengChen2023}. 
Specifically, these numerical studies find that strong SC emerges from lighter electron doping rather close to half filling and extends over a much broader range than those typically observed experimentally.
Yet, a consensus has emerged in the community on two aspects of electron-doped cuprates: 
consistent findings on superconductivity from unbiased numerical studies of pure single-band models \cite{ShoushuGong2021, ShengtaoJiang2021, ShengtaoJiang2022, XinLu2023,HaoXu2023,FengChen2023} and the adequacy of a single-band model to describe the real materials, where doping involves only the $\mathrm{d}_{x^2-y^2}$ orbital of the copper. These agreements suggest that resolving the discrepancies in electron-doped cuprates is more computationally manageable, theoretically feasible, and of critical importance for understanding the underlying physics.
Such an exploration calls for the study of single-band Hubbard or $t$-$J$ models with additional ingredients that accurately and simultaneously capture the essential physics of the corresponding materials~\cite{Takagi1989, Uefuji2001, Motoyama2007, Armitage2010, Scalapino2012}, including robust AF, weak charge density wave (CDW) and absent SC at light doping, coexistence of AF and SC with increasing doping, narrow SC region around optimal doping, and the existence of pseudogap (PG).

In one- and two-dimensional materials, it is noticeable that Coulomb screening is weaker than that in three dimensions, making it challenging to restrict the inter-site interactions between electrons to on-site interactions only.  
Recent studies have accumulated evidence demonstrating the crucial role of inter-site  repulsion in both hole- and electron-doped cuprates \cite{Sau2014, Misawa2014, ZuoDongYu2017,Hirayama2018, Hirayama2019,Zinni2021,Boschini2021, Bejas2022, Banerjee2022,Scott2023,Riegler2023}. 
These findings suggest that nearest-neighbor electron repulsion is pertinent in understanding the static and dynamic charge order \cite{Boschini2021, Bejas2022, Scott2023} as well as its interplay with spin order \cite{Riegler2023}, consistent with experiments of compounds $\mathrm{Nd}_{2-x} \mathrm{Ce}_x \mathrm{CuO}_4$ (NCCO) \cite{Lee2014,Neto2016,Neto2018,Hepting2018}. 
Meanwhile, significant attention has also been given to inter-site attraction such as phonon effect~\cite{ZhuoyuChen2021,YaoWang2021,ZhongBingHuang2021,MiJiang2022, DaiWeiQu2022,HaoXinWang2022, HaoXinWang2022SSH, XunCai2023,ChengPeng2023,TaTang2023,ZongshengZhou2023}.

\begin{figure}[tbp]
\begin{center}
\includegraphics[width=0.4\textwidth]{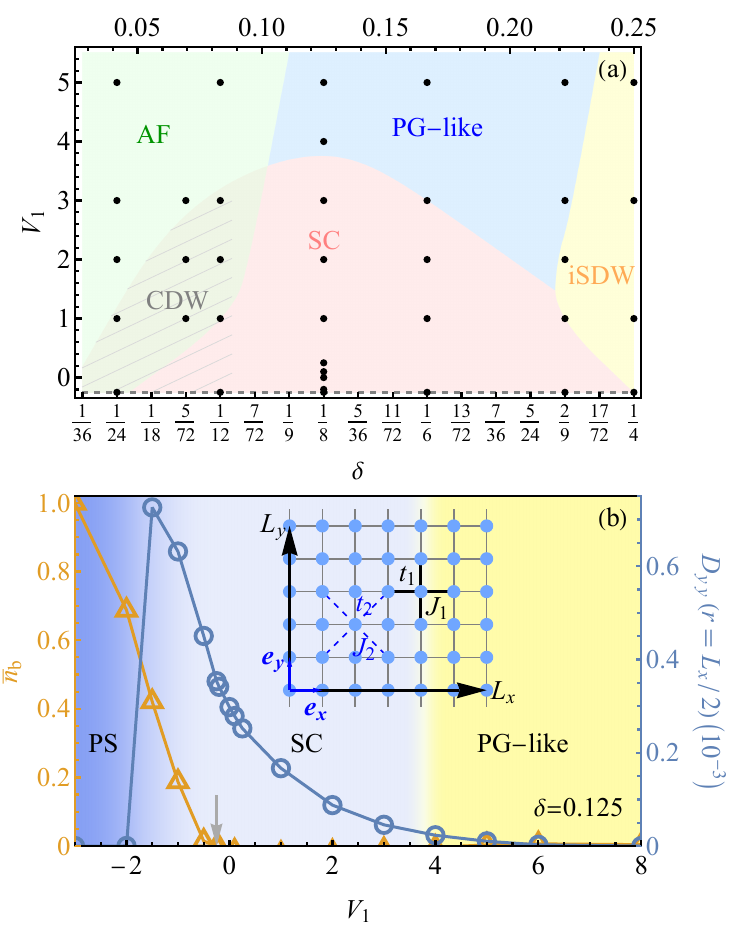}
\end{center}
\par
\renewcommand{\figurename}{Fig.}
\caption{Distinct phases in the $t$-$J$-$V$ model.   
(a) Phase diagram as a function of nearest-neighbor interaction $V_1$ and  dopings $\delta$. There are Néel antiferromagnetic order (AF), pseudogap-like (PG-like), superconductivity (SC), incommensurate spin density wave (iSDW), and charge density wave (CDW) phases, as well as regions where some of these phases coexist.  Black dots are the parameters we calculate.  
(b) Fixing $\delta=1/8$, three distinct phases as a function of $V_1$: phase separation (PS), SC, and PG-like phase. The yellow triangles and blue circles denote $\bar{n}_b$ [see Eq.S5 in supplementary~\cite{SM}], a quantity characterizing PS, and $D_{yy}(r=L_x/2)$, the pair correlations at $r=L_x/2$, respectively. Other choices of $r$ exhibit similar behavior~\cite{SM}. 
The inset depicts the square lattice and model parameters.
The gray dashed line in (a) and the gray arrow in (b) correspond to the standard 
$t$-$J$ model with $V_{ij}=-0.25J_{ij}$.The energy unit is chosen as $J_1$.
Here, $t_2/t_1=0.2$ corresponds to electron-doped {cuprates} and both {phase} diagrams are identified on $N=24\times6$ systems.
}
\label{Fig_PhaseDiagram}
\end{figure}

Motivated by the above, we examine the roles of the inter-site interactions in the $t$-$J$-$V$ model on the square lattice, aiming at identifying a minimal model capable of describing the phase diagram of the electron-doped cuprates, as validated by experiments.  
The Hamiltonian 
is given by
\begin{equation}\label{eq:H}
\begin{split}
 H=&-\sum_{\{ij\}, \sigma} t_{ij}\mathcal{P}\left(\hat{c}_{i, \sigma}^{\dagger} \hat{c}_{j, \sigma}+\text { H.c. }\right)\mathcal{P}\\
 &+\sum_{\{i j\}} J_{i j}{\mathbf{S}}_i \cdot {\mathbf{S}}_j+\sum_{\{ij\}}V_{ij}n_in_j.
 \end{split}
\end{equation}
Here, $ {c}_{i, \sigma}^{\dagger}$ and $  {c}_{i, \sigma}$ are the electron creation and annihilation operators with spin-$\sigma$ 
at site $i$. $  {\mathbf{S}}_i$ is the spin-$1 / 2$ operator and  $ {n}_i$ 
represents electron number. $\mathcal{P}$ projects to the single-occupancy subspace. We consider both nearest-neighbor (NN) and next-nearest-neighbor (NNN) bonds, the hopping amplitude $t_{ij}$ equates to $t_1$ for NN and $t_2$ for NNN, with the corresponding values for $J_{ij}$ and $V_{ij}$ being $J_1$, $J_2$, and $V_1$, $V_2$, as illustrated in the inset of Fig.~\ref{Fig_PhaseDiagram}(b). The energy unit is $J_1$.
Specifically, we choose parameters $t_1 / J_1=3.0, J_2 / J_1=\left(t_2 / t_1\right)^2$ with ratio $t_2 / t_1 = 0.2$ and $V_2/V_1=J_2/J_1$, corresponding to the parameter region of {electron-doped} cuprates \cite{Pavarini2001,Tanaka2004,Kim1998}. 
We investigate the ground-state properties by density matrix renormalization group (DMRG) \cite{Whitedmrg,DMRG3,DMRG2}. The square-lattice size is $N = L_x \times L_y$, where $L_x$ and $L_y$ represent the cylinder length and circumference, respectively. Considering the varying convergence rates at different parameters, we set the bond dimension up to $D=45000$ in U(1)$\times$U(1) DMRG and $D=18000\sim 24000$ in U(1)$\times$SU(2) DMRG [see supplementary~\cite{SM} for DMRG details].

\begin{figure}[tpb]
\begin{center}
\includegraphics[width=0.4\textwidth]{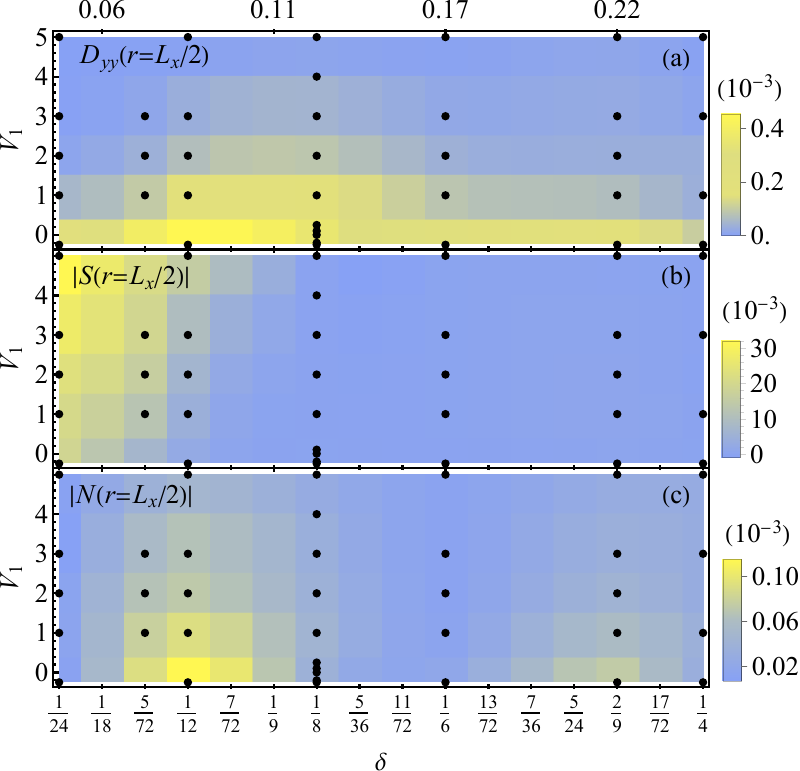}
\end{center}
\par
\renewcommand{\figurename}{Fig.}
\caption{Effect of nearest-neighbor interaction$V_1$ on SC, AF and CDW phases with respect to doping $\delta$. Panels (a-c) plot the amplitudes of pair correlations $D_{yy}(r)$, 
spin correlations $S(r)$,  
and charge density correlations $N(r)$ at $r=L_x/2$.
The amplitudes of $|S(r=L_x/2)|$ around $\delta\sim1/4$  are significantly smaller than those with $\delta\lesssim1/12$, thereby masking the signature of iSDW in panel (b) [see supplementary~\cite{SM}]. 
Black dots are calculated parameters for $N=24\times6$. 
}
\label{Fig_DeltaVCor}
\end{figure}

\emph{Phase diagram and main findings.---} 
Based on the DMRG simulation of the $t$-$J$-$V$ model, we identify different phases and construct the ground-state phase diagram as a function of inter-site interactions $V$ and doping concentrations $\delta$, as depicted in Fig.~\ref{Fig_PhaseDiagram}(a). 
Our findings indicate that the $t$-$J$-$V$ model, with inter-site repulsion $2J_1 \lesssim V_1 \lesssim 3J_1$ comparable to the hopping amplitude $t_1$, more accurately reflects the experimental observations of electron-doped cuprates~\cite{Armitage2010,Neto2016} than the conventional Hubbard and $t$-$J$ models, especially in terms of SC, AF, PG, and CDW phases. This suggests the minimal model of electron-doped cuprates is the $t$-$J$-$V$ model.

\begin{figure}[tpb]
\begin{center}
\includegraphics[width=0.48\textwidth]{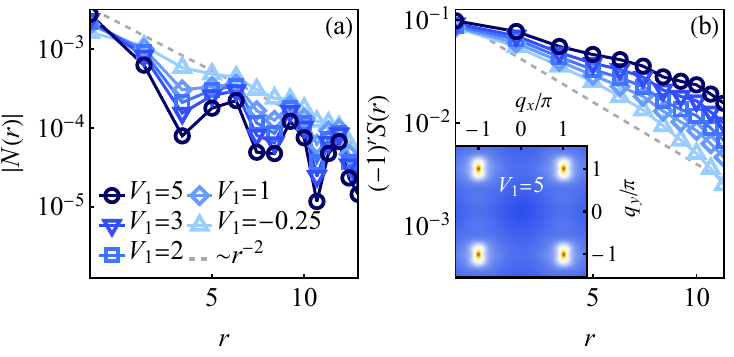}
\end{center}
\par
\renewcommand{\figurename}{Fig.}
\caption{
Effect of nearest-neighbor interaction $V_1$ on charge density correlations in panel (a) and spin correlations in panel (b) for $\delta=1/12$ and $N=24\times6$. Panels (a,b) are labeled identically. The inset of (b) shows a contour plot of static spin structure factor $S(\mathbf{q})$. 
}
\label{Fig_CDWSDWDelta112}
\end{figure}

Specifically, within $2 \lesssim V_1/J_1 \lesssim 3$, SC is absent and AF dominates at light doping levels; AF coexists with SC and weak CDW as doping increases;
around optimal doping, SC becomes dominant, but is confined to a narrower doping range compared to the much broader range observed in the $t$-$J$ limit [dashed gray line in Fig.~\ref{Fig_PhaseDiagram}(a)]; a tentative incommensurate {spin density wave} (iSDW) phase is predicted to emerge at even larger doping~\cite{SM}.

Moreover, we reveal the impact of inter-site interactions on various phases, as shown in Figs.~\ref{Fig_PhaseDiagram},~\ref{Fig_DeltaVCor},~\ref{Fig_CDWSDWDelta112},~\ref{Fig_CorVSV}. At light doping, inter-site interactions predominantly favor AF [see Fig.~\ref{Fig_DeltaVCor}(b) and Fig.~\ref{Fig_CDWSDWDelta112}(b)] and suppress both SC and CDW quasi-long-range orders [see Fig.~\ref{Fig_DeltaVCor}(a,c) and Fig.~\ref{Fig_CDWSDWDelta112}(a)]. 
Around optimal doping, these interactions preferentially support a pseudogap-like (PG-like) phase while suppressing SC [see Fig.~\ref{Fig_PhaseDiagram}, Fig.~\ref{Fig_DeltaVCor}(a) and Fig.~\ref{Fig_CorVSV}(a)]. At higher doping around $\delta\sim1/4$, the impact of inter-site interactions becomes less pronounced [see Fig.~\ref{Fig_DeltaVCor}].

In particular, we 
especially focus on SC region and elucidate the role of inter-site interactions around optimal doping.
As shown in Fig.~\ref{Fig_PhaseDiagram}(b), when the inter-site interaction shifts from attraction to repulsion, 
the system undergoes successive phase transitions: from phase separation for large inter-site attraction to uniform $d$-wave SC, then to a PG-like phase with strong CDW, SDW and superconducting fluctuations. 
Within the SC phase, when shifting the inter-site interaction from attraction to repulsion, a notable suppression of SC is observed until it is ultimately destroyed at larger repulsion. Such numerical observation has also been microscopically interpreted by slave-boson mean-field analysis. 
Remarkably, our finding of the phase separation under large inter-site attraction [see supplementary~\cite{SM}] implies that the effective inter-site interactions could be repulsive, 
since
there is no clear experimental evidence of the phase separation in the electron-doped cuprates \cite{Harima2001,Damascelli2003,Armitage2010}.
This finding also supports the  $t$-$J$-$V$ model with significant inter-site repulsion as the minimal model of the electron-doped cuprates.

\begin{figure}[!t]
\begin{center}
\includegraphics[width=0.48\textwidth]{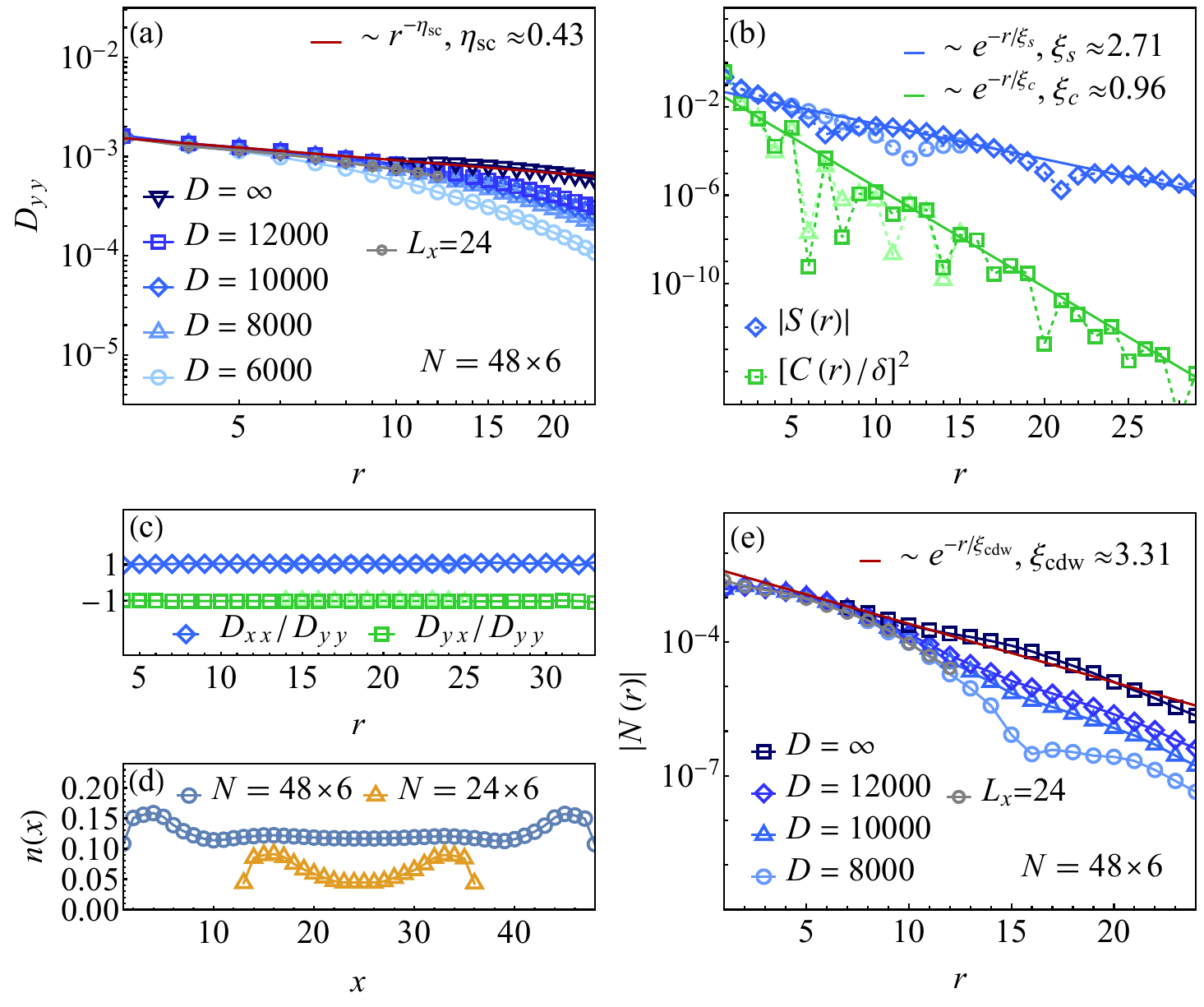}
\end{center}
\par
\renewcommand{\figurename}{Fig.}
\caption{Properties of SC phase. (a) Pair-pair correlations.
(b) spin correlations and renormalized single-particle propagator.
(c) The ratio of pair correlations between different directional bonds. Here, data of $L_x=24$ are offset horizontally for clarity, and darker and lighter colors in panels (b,c) imply $L_x=48$ and $L_x=24$, respectively.
(d) The charge density distribution. Here, $L_x=24$ data are offset horizontally and vertically for clarity.
(e) The charge density correlations.
A second-order polynomial fitting of $1/D$ is used to
extrapolate bond dimension $D=\infty$ for $D_{yy}$ (a) and $N(r)$ (e).
Here, we consider the typical parameter $V_1=-1$ 
for $N=24\times6$ and $N=48\times6$ 
with $\delta=1/8$. 
}
\label{Fig_SC_V-1}
\end{figure}


\emph{AF and CDW with varying inter-site interactions at light doping.---}
At light doping, the amplitudes of spin correlations $S(r)\equiv \langle \mathbf{S}_{{\mathbf{r}}_0}\cdot \mathbf{S}_{{\mathbf{r}_0+r\boldsymbol{e}_x}}\rangle$ [see Fig.~\ref{Fig_DeltaVCor}(b)] and charge density correlations $N(r)\equiv\langle n_{\mathbf{r}_0}n_{{\mathbf{r}_0+r\boldsymbol{e}_x}}\rangle-\langle n_{\mathbf{r}_0}\rangle\langle n_{{\mathbf{r}_0+r\boldsymbol{e}_x}}\rangle$ [see Fig.~\ref{Fig_DeltaVCor}(c)] at bulk of the system with $r=L_x/2$ indicate that inter-site interactions mainly favor SDW and suppress CDW.
To provide a specific illustration of this trend, we depict   $|N(r)|$ and $|S(r)|$ for $\delta=1/12$ with various inter-site interactions in Fig.~\ref{Fig_CDWSDWDelta112}. 
In the limit of the {standard $t$-$J$ model}, i.e., $V_1=-0.25$, the CDW slightly dominates over SDW, since $|N(r)|$ and $|S(r)|$ decay at rates comparable to and faster than $\sim r^{-2}$, respectively. However, with the increase of inter-site interactions, $|N(r)|$ decays more rapidly, while $|S(r)|$ shows a slower decay, becoming  power-law behavior at large inter-site repulsion, thereby demonstrating suppression of CDW and a preference for SDW at light doping. 
We identify the AF at light doping by demonstrating that $S(r)$ exhibits the AF property $(-1)^rS(r)>0$ [see Fig.~\ref{Fig_CDWSDWDelta112}(b)] and the peaks of the  static spin structure factor $S(\mathbf{q})\equiv\sum_{\mathbf{i},\mathbf{j}} 
\langle {\mathbf{S}_\mathbf{i}\cdot \mathbf {S}_\mathbf{j}} \rangle 
e^{i \mathbf{q}\cdot (\mathbf{i}-\mathbf{j})}/N$ stabilize at the momentum $\mathbf{q}_0=(\pm\pi,\pm\pi)$, as shown in supplementary \cite{SM} and the inset of Fig.~\ref{Fig_CDWSDWDelta112}(b).
While such preference for AF by inter-site repulsion is intriguing for theoretical study, the suppression of the CDW even appears to be counterintuitive.

\emph{Superconductivity versus inter-site interactions around optimal doping.---} {We 
take $\delta=1/8$ as an example and study the impact of inter-site interactions on superconductivity around optimal doping. 
Similar observations are also noted at $\delta=1/6$.
}
We examine the pair correlations defined by
\begin{equation}\label{eq:D}
{\footnotesize
D_{\alpha\beta}(\mathbf{r}) \equiv \left\langle \hat{\Delta}^{\dagger}_{\alpha}(\mathbf{r}_0) \hat{\Delta}_{\beta}(\mathbf{r}_0 +\mathbf{r})\right\rangle,
}
\end{equation}
where the pair operator $\hat{\Delta}_{\alpha}(\mathbf{r}) \equiv \frac 1 {\sqrt{2}}\sum_{\sigma}\sigma {c}_ {\mathbf{r}, \sigma} {c}_ {\mathbf{r}+\boldsymbol{e}_\alpha, -\sigma}$, and $\alpha, \beta = x, y$. 
In quasi-one-dimensional cylinders, we explore the presence of quasi-long-range order characterized by $D_{\alpha\beta}({r})\sim r^{-\eta_{\mathrm{sc}}}$. 
Specifically, $\eta_{\mathrm{sc}} < 2$ indicates a divergent superconducting susceptibility in two dimensions as the temperature $T\to 0$.

For a representative $V_1=-1$ in the SC phase [see Fig.~\ref{Fig_SC_V-1}(a)], the pair correlations exhibit a slow power-law decayed rate $\eta_\mathrm{sc}\approx0.43$.
Moreover, we find exponentially decaying spin correlations $S(r)$ 
and single-particle propagators $C(r)=\sum_{\sigma}\langle c_{\mathbf{r}_0,\sigma}^\dagger c_{{\mathbf{r}_0+r\boldsymbol{e}_x},\sigma}\rangle $, as illustrated in Fig.~\ref{Fig_SC_V-1}(b).
These results suggest robust SC. 
We further 
find that $D_{yx}$ is always negative, while $D_{yy}$ and $D_{xx}$ are positive. The ratio $D_{y x}/D_{y y}$ is close to $-1$ while $D_{xx}/D_{y y}\approx1$ [see Fig.~\ref{Fig_SC_V-1}(c)], suggesting the equal amplitude but opposite signs for the pairs between $x$ and $y$ bonds, consistent with $d$-wave pairing symmetry.

In the SC phase, we also examine the charge density distribution $\langle \hat{n}(x,y)\rangle$. Due to the translational invariant along the $\boldsymbol{e}_y$ direction, we focus on the distribution along $\boldsymbol{e}_x$ and define ${n}(x) \equiv {\sum_{y} \langle \hat{n}(x,y)\rangle/L_{y}}$, where $\hat{n}(x,y)$ is the density operator for the doped charge at site $(x,y)$.
The uniform profile of ${n}(x)$, which remains robust across different system sizes, is depicted in Fig.\ref{Fig_SC_V-1}(d).
This is consistent with the rapid decay of charge density correlations $N(r)$ 
in the SC phase.
Notably, exponential decay in $N(r)$ emerges when considering a relatively larger inter-site attraction [see the semi-logarithmic plot in Fig.\ref{Fig_SC_V-1}(e)]. 
This contrasts with the power-law decay of $N(r)$ in the SC phase of $t$-$J$ model~\cite{ShoushuGong2021}. This difference in the decay behavior of $N(r)$ is a distinct SC feature resulting from inter-site attraction.

\begin{figure}[!t]
\begin{center}
\includegraphics[width=0.45\textwidth]{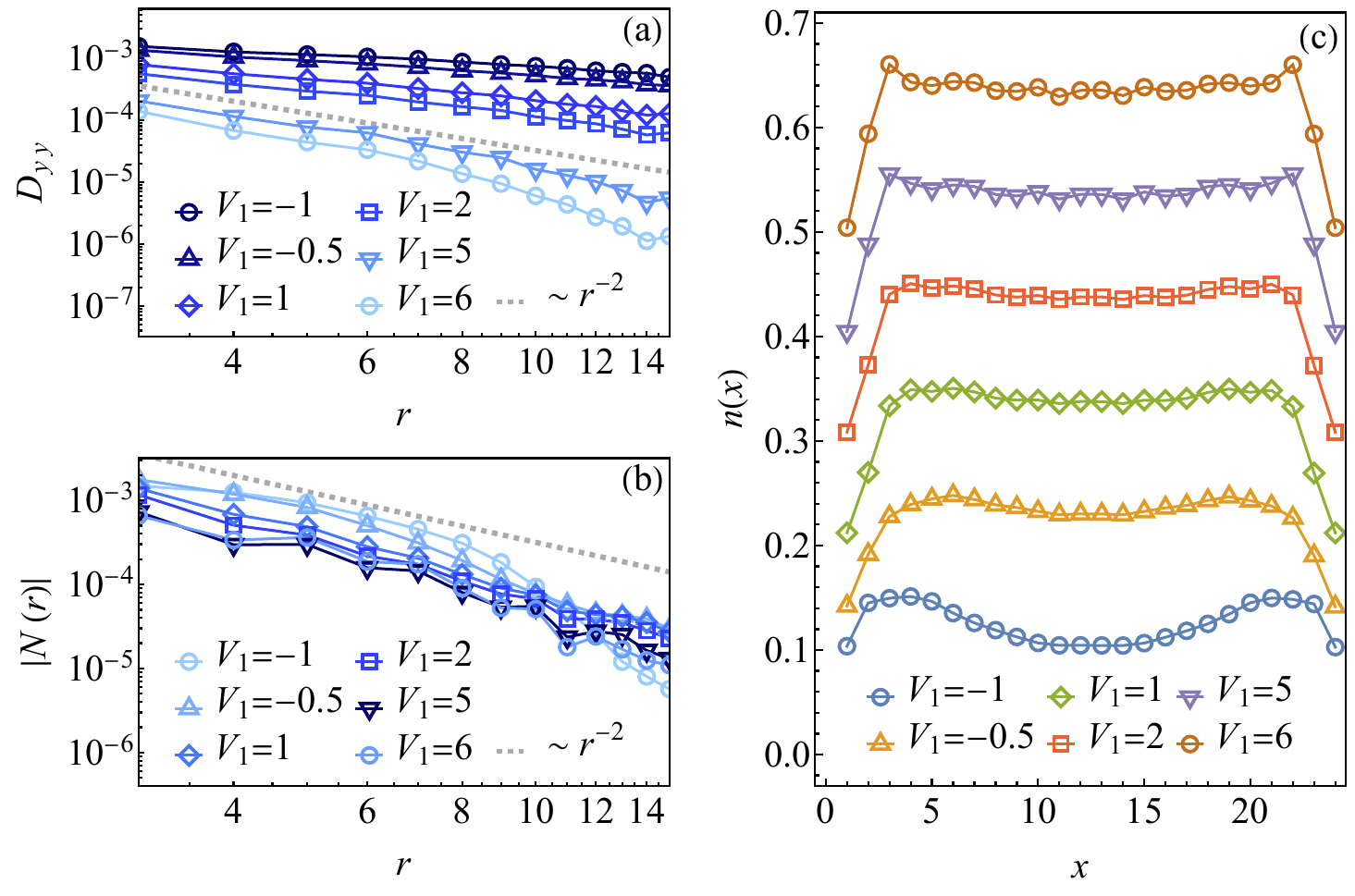}
\end{center}
\par
\renewcommand{\figurename}{Fig.}
\caption{Properties of the system with tuning $V_1$.
Panels (a,b) show pair correlations (a) and charge density correlations (b).
Here the dashed gray line signifies $r^{-2}$. Darker colors imply a slower decay rate.
(c) The charge density distribution $n(x)$. Data are offset vertically for clarity.
Here, we consider $\delta=1/8$ on $N=24\times6$ system.
}
\label{Fig_CorVSV}
\end{figure}

We further calculate the pair correlations across a range of $V_1$ from inter-site attraction to repulsion, and find a weakening of the pair correlations with increasing $V_1$, as shown in Fig.~\ref{Fig_PhaseDiagram}(b) and Fig.~\ref{Fig_CorVSV}(a). Despite this, the charge density distribution $n(x)$ always exhibits a uniform profile 
across the broad $V_1$ range [see Fig.~\ref{Fig_CorVSV}(c)].
Such uniform charge density distribution supports the translation invariance implemented in our slave-boson mean-field approach, with which we examine the trend of SC with respect to $V_1$. Specifically, using slave-boson representation, 
the mean-field Hamiltonian [see supplementary \cite{SM}] can be derived as  
\begin{equation*}
\begin{split}
H^\mathrm{MF}=&-B\sum_{\{ ij \},\sigma}t_{ij}{( {f_{i\sigma}^{\dagger}f_{j\sigma}} +h.c. )}-\chi \sum_{\{ ij \}}t_{ij}{( b_{i}^{\dagger}b_j +h.c.)}\\
&-\frac{\chi}{4}  \sum_{\{ ij \},\sigma}(J_{ij}+2V_{ij}){( {f_{i\sigma}^{\dagger}f_{j\sigma}} +h.c.)}\\
&-\frac{1}{2}\sum_{\{ ij \}}( J_{ij}-V_{ij} ) {[ \Delta ^{\ast}( f_{j\uparrow}f_{i\downarrow}-f_{j\downarrow}f_{i\uparrow} ) +h.c. ]}\\
&-\mu \sum_{i,\sigma}{f_{i\sigma}^{\dagger}f_{i\sigma}}+\lambda \sum_i{( \sum_{\sigma}{f_{i\sigma}^{\dagger}f_{i\sigma}} +b_{i}^{\dagger}b_i-1 )}
\end{split}
\end{equation*}
where $B=\langle b_ib_{j}^{\dagger} \rangle$, $\chi =\sum_{\sigma}{\langle f_{i\sigma}^{\dagger}f_{j\sigma} \rangle}$, and $\Delta=\langle f_{j\uparrow}f_{i\downarrow}-f_{j\downarrow}f_{i\uparrow} \rangle $. Here, $\mu$ is the chemical potential, and $\lambda$ is the Lagrange multiplier. Notably, we decouple the interaction term $n_in_j$ into pairing and hopping channels, instead of replacing it by a constant $(1-\delta)^2$ as usual practice. 
Consequently, 
the signs of $J_{ij}$ and $V_{ij}$ are opposite in the pairing channel. This indicates that repulsive $V_{ij}$ is energetically unfavorable for pairing, contrasting with the enhancing impact of attractive $V_{ij}$.
Further examination through solving the self-consistent equations reveals that 
the critical temperature $T_c$ decreases when the inter-site interaction shifts from attraction to repulsion \cite{SM}. These mean-field analyses are consistent with numerical observations.

For inter-site repulsion, $N(r)$ is relatively insensitive to $V_1$, contrasting the behavior of  pair correlations $D_{yy}$ [see Fig.\ref{Fig_CorVSV}(a)], and follows more closely a power-law form, $\sim r^{-\eta_\mathrm{cdw}}$, with $\eta_\mathrm{cdw}\gtrsim 2$, as illustrated in Fig.~\ref{Fig_CorVSV}(b).
This trend resembles experimental findings in electron-doped cuprates, where the response of the charge order to SC is relatively less sensitive near optimal doping~\cite{Neto2016}.
Nonetheless, our numerical results still reveal an observable relationship: the decay of $N(r)$ occurs at a slightly slower rate during the suppression of SC within the SC regime.
This hints at a subtle competition between the CDW and the SC modulated by inter-site interactions.

\begin{figure}[!t]
\begin{center}
\includegraphics[width=0.45\textwidth]{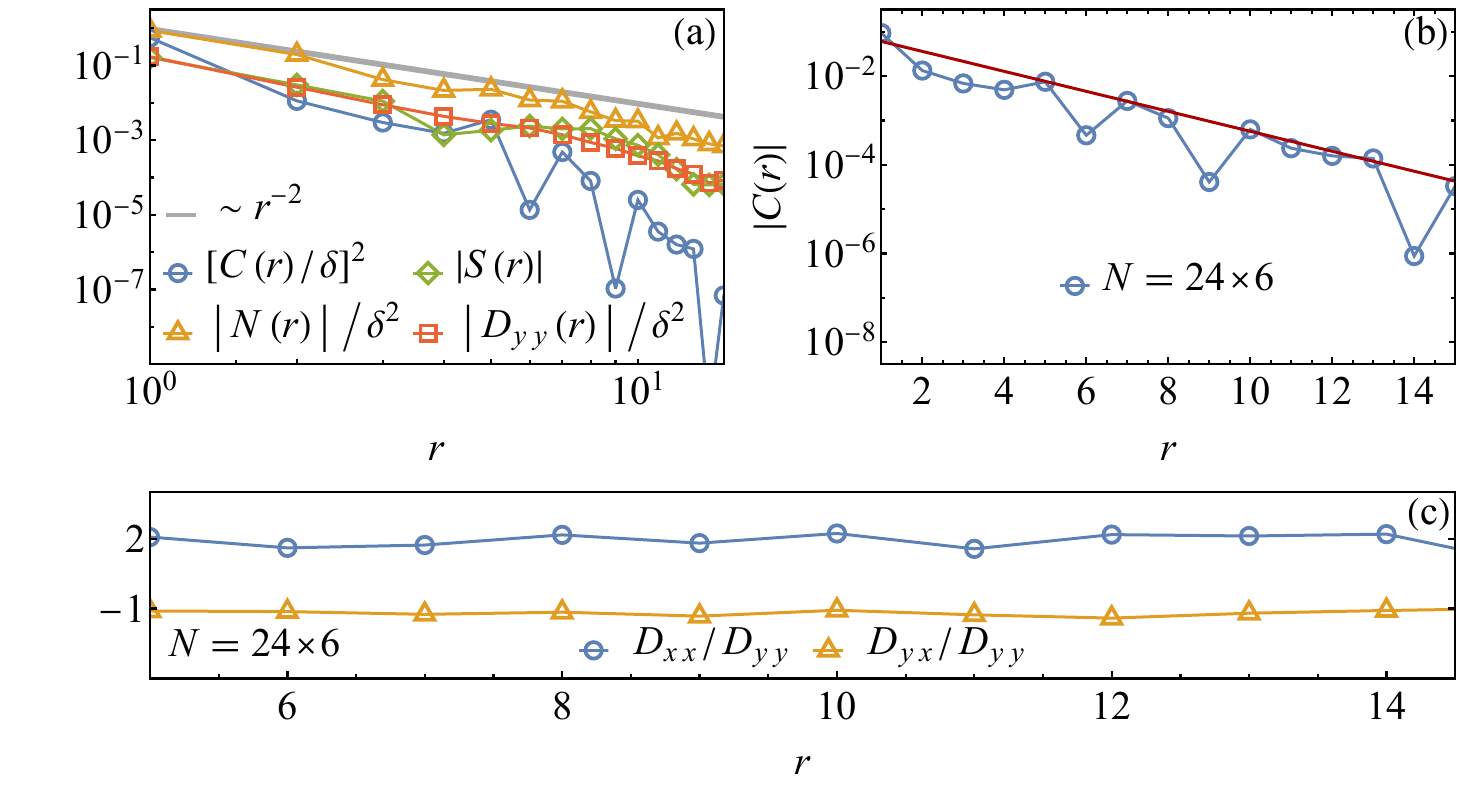}
\end{center}
\par
\renewcommand{\figurename}{Fig.}
\caption{Properties of pseudogap-like (PG-like) phase.
(a) Various correlations. We renormalize different correlations to facilitate a direct comparison.
(b) Exponentially decayed single particle propagator $|C(r)|$.
(c) The ratio of pair correlations between different directional bonds.
Here, we consider a typical parameter $V_1=6$ on a $N=24\times6$ system with $\delta=1/8$. 
}
\label{Fig_PG}
\end{figure}

\emph{Pseudogap-like phase for large inter-site repulsion around optimal doping.---} Around the optical doping, with increasing $V_{ij}$, the pair correlation $D_{yy}$ experiences suppression and exhibits a significantly faster decay rate $\eta\gtrsim2$ than that observed in the SC phase, as illustrated by the renormalized quantity $ |D_{yy}(r)|/\delta^2$ in Fig.~\ref{Fig_PG}(a).
Consequently, a PG-like phase emerges, characterized by the coexistence of strong CDW, superconducting, and SDW fluctuations. The intricate interplay of these fluctuations is evident in Fig.~\ref{Fig_PG}(a), where the renormalized $|N(r)|/\delta^2$ exhibits a slight prominent amplitude over $|S(r)|$ and $|D_{yy}(r)|/\delta^2$,  signifying the marginally predominant role of CDW fluctuations over the SDW and superconducting fluctuations.
Notably, in Fig.~\ref{Fig_PG}(b), the single particle propagator $|C(r)|$ decays exponentially, suggesting the charge insulation along $\boldsymbol{e}_x$.
Additionally, as illustrated in Fig.~\ref{Fig_PG}(c), we find the ratios $D_{y x}/D_{y y}\approx-1$ remain valid, though the departure of $D_{xx}/D_{y y}$ from unity is more noticeable. This suggests the local $d$-wave pairing symmetry remains relatively robust in this PG-like phase even without quasi-long-range SC order.



\emph{Summary.---} Our investigation of the $t$-$J$-$V$ model on the square lattice identifies distinct correlated phases and establishes a ground-state phase diagram as a function of inter-site interactions and doping concentrations [see Fig.~\ref{Fig_PhaseDiagram}]. These findings suggest the minimal model of electron-doped cuprates should be the $t$-$J$-$V$ model with $2 \lesssim V_1/J_1 \lesssim 3$ instead of the conventional $t$-$J$ models. Notably, the putative inter-site repulsion $V_1$ is comparable to the hopping amplitude $t_1$. 
Furthermore, we elucidate the role of inter-site interactions at various doping levels: at light doping levels, these interactions favor N\'{e}el antiferromagnetic order and inhibit CDW; around optimal doping levels, they promote a PG-like phase while suppressing SC; at higher doping levels, their impact becomes less significant, while a tentative iSDW phase is predicted to emerge.
We also illustrate the role of inter-site interactions in SC region around optimal doping, and reveal the successive phases from phase separation to uniform SC, and further to a PG-like phase with competition among CDW, SC, and SDW.  Therefore, our study provides insights relevant to understanding electron-doped cuprates materials~\cite{Armitage2010}. Moreover, our work may also stimulate future studies on the  $t$-$J$-$V$-type models applied to other lattice geometries, such as the triangular lattice~\cite{Zhu2022Doped,Chen2022Proposal, Jiang2021Superconductivity, KevinHuang2022, Huang2023Quantum, ZhengZhu2023}, where inter-site repulsion manifests prominently in materials like transition-metal-dichalcogenide bilayers.

\begin{acknowledgments}

Q. C. is supported by the Fundamental Research Funds for the Central Universities.
F. Z. acknowledges the support of the National Natural Science Foundation of China (Grant No.11920101005), the Strategic Priority Program of CAS  (Grant No. XDB28000000), and the Ministry of Science and Technology (Grant No. 2022YFA1403900) by Chinese Academy of Sciences under contract No. JZHKYPT-2021-8, Innovation program for Quantum Science and Technology (Grant No. 2021ZD0302500).
Z. Z. acknowledges the support of the National Natural Science Foundation of China (Grant No.12074375), the Fundamental Research Funds for the Central Universities and the Strategic Priority Research Program of CAS (Grant No.XDB33000000).
\end{acknowledgments}

\onecolumngrid

\newpage

\renewcommand{\theequation}{S\arabic{equation}}
\setcounter{equation}{0}
\renewcommand{\thefigure}{S\arabic{figure}}
\setcounter{figure}{0}
\renewcommand{\bibnumfmt}[1]{[S#1]}

\section{Supplementary Materials for\\``{Phase Diagram of the Square-Lattice $t$-$J$-$V$ Model for Electron-Doped Cuprates
}"}

\section{DMRG details and Correlation functions}

We investigate the ground-state properties by density matrix renormalization group (DMRG).
The square lattice is defined by the primitive vectors ${\boldsymbol{e}_{{x}}}=(1,0)$, $\boldsymbol{e}_{{y}}=(0,1)$ and is wrapped on cylinders with a lattice spacing of unity, as shown in the inset of Fig.1(b). The system size is denoted as $N = L_x \times L_y$, where $L_x$ and $L_y$ correspond to the cylinder length and circumference, respectively. The doping concentration is represented by $\delta = N_0 / N$, with $N_0$ indicating the number of doped charges. In our study, we primarily focus on the width-6 cylinders, i.e., $L_y=6$. Considering the varying convergence rates at different parameters, we set the bond dimension $D$ up to $D\approx 45,000$ when implementing symmetries with $U(1)$ charge and $U(1)$ spin-$S_z$, and $D=18,000$  (or $D=24,000$ in specific cases)  when implementing symmetries with $U(1)$ charge and $SU(2)$ spin, both give consistent results.

To facilitate a direct comparison of various correlations at finite doping $\delta$, we summarize the definitions of the renormalized correlations as following
\begin{itemize}
  \item the renormalized single particle propagator $\mathrm{CC}(r)\equiv [C(r)/\delta]^2$, where the single particle propagator $C(r)$ is
      \begin{equation}\label{}
        C(r)=\sum_{\sigma}\langle c_{\mathbf{r}_0,\sigma}^\dagger c_{{\mathbf{r}_0+r\boldsymbol{e}_x},\sigma}\rangle .
      \end{equation}
  \item the renormalized spin correlations $\mathrm{SS}(r)\equiv |S(r)|$, where the spin correlations $S(r)$ is
      \begin{equation}\label{eqS:Sr}
        S(r)\equiv \langle \mathbf{S}_{{\mathbf{r}}_0}\cdot \mathbf{S}_{{\mathbf{r}_0+r\boldsymbol{e}_x}}\rangle.
      \end{equation}
  \item the renormalized pair correlations  $\mathrm{DD}(r)\equiv |D_{yy}(r)|/\delta^2$, where the pair correlations $D_{\alpha\beta}({r})$ is
      \begin{equation}\label{eqS:D}
        D_{\alpha\beta}({r}) \equiv \left\langle \hat{\Delta}^{\dagger}_{\alpha}(\mathbf{r}_0) \hat{\Delta}_{\beta}(\mathbf{r}_0 +{r}\boldsymbol{e}_x)\right\rangle,\quad \text{with }
        \hat{\Delta}_{\alpha}^\dagger(\mathbf{r}_0) \equiv \frac 1 {\sqrt{2}}\left(c_{\mathbf{r}_0+\boldsymbol{e}_{\alpha}, \downarrow}^{\dagger} c_{\mathbf{r}_0, \uparrow}^{\dagger}-c_{\mathbf{r}_0+\boldsymbol{e}_{\alpha}, \uparrow}^{\dagger} c_{\mathbf{r}_0, \downarrow}^{\dagger}\right).
      \end{equation}
      Here, $\alpha, \beta$ denote the bonds $\alpha, \beta = x, y$.
  \item the renormalized charge-density correlations $\mathrm{NN(r)}\equiv |N(r)|/\delta^2$, where the charge-density correlation $N(r)$ is
      \begin{equation}\label{}
        N(r)\equiv\langle n_{\mathbf{r}_0}n_{{\mathbf{r}_0+r\boldsymbol{e}_x}}\rangle-\langle n_{\mathbf{r}_0}\rangle\langle n_{{\mathbf{r}_0+r\boldsymbol{e}_x}}\rangle .
      \end{equation}
\end{itemize}
Unless otherwise stated, we set the reference position $\mathbf{r}_0=(L_x/4,y_0)$ for an unbiased calculation to avoid the boundary effect. Due to the translational invariance of our geometry along the $\boldsymbol{e}_y$ direction, the results are not influenced by the value of $y_0$.

\section{Superconductivity with varying inter-site interactions}
\begin{figure}[H]
\begin{center}
\includegraphics[width=0.3 \textwidth]{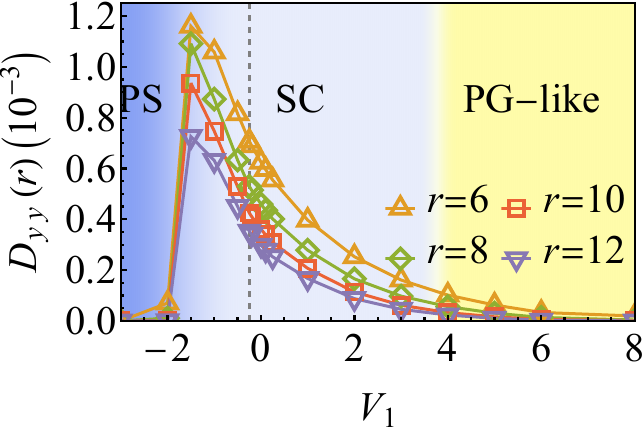}
\end{center}
\par
\renewcommand{\figurename}{Fig.}
\caption{Quantum phase diagram of the $t$-$J$-$V$ model with $\delta=1/8$. We identify three distinct phases (PS for phase separation, SC for superconductivity, and PG-like for pseudogap-like phase) of the extended $t$-$J$-$V$ model as a function of $V_1$.
The dashed gray line corresponds to the $t$-$J$ limit with $V_1=-0.25$, which implies the absence of the inter-site electron interaction. The phase diagram is determined with $L_x=24$ and $L_y=6$.
}
\label{FigS_PhaseDiagramSC}
\end{figure}

\begin{figure}[H]
\begin{center}
\includegraphics[width=0.58\textwidth]{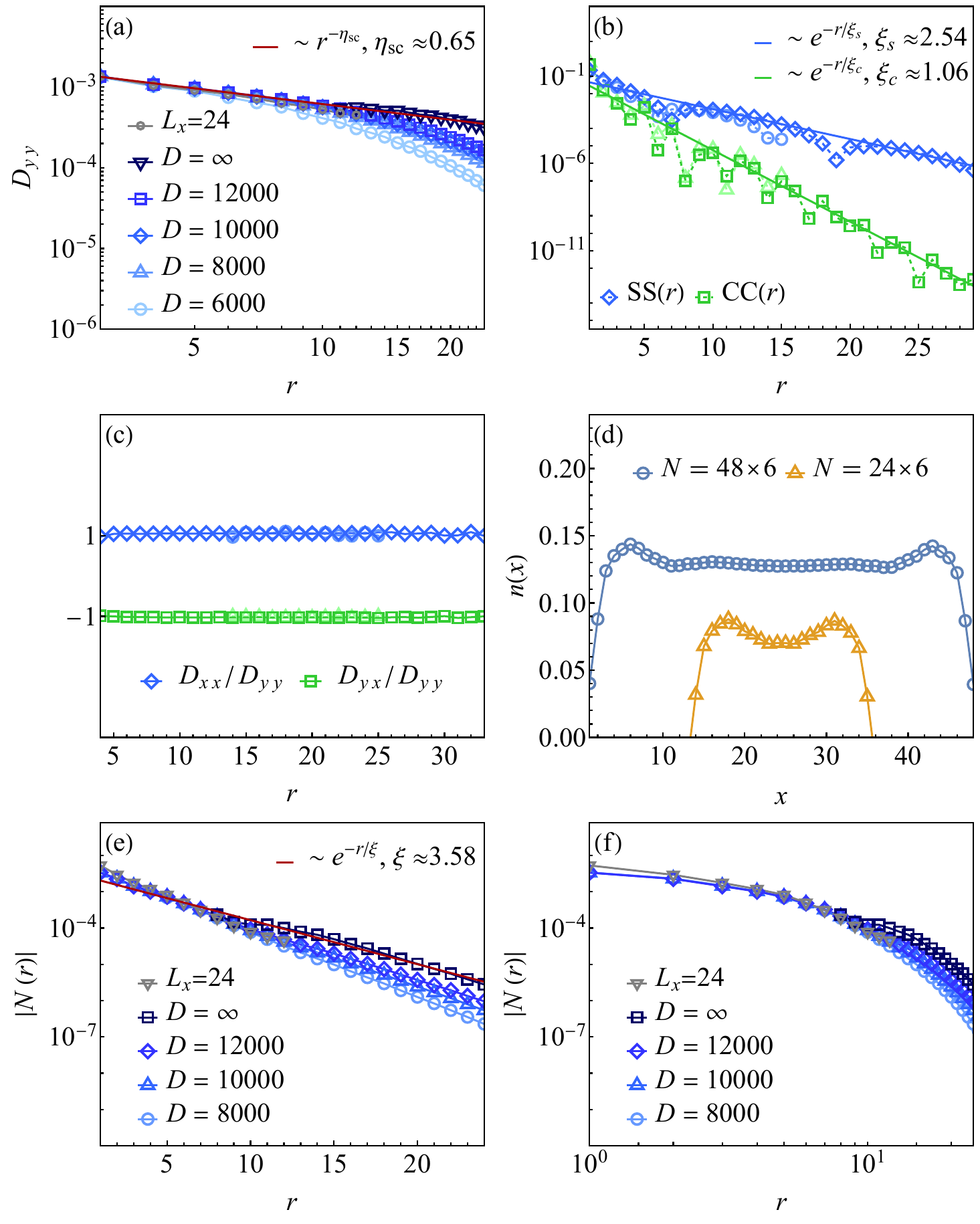}
\end{center}
\par
\renewcommand{\figurename}{Fig.}
\caption{SC properties at typical parameter $V_1=-0.5$ of the phase for $N=24\times6$ and $N=48\times6$ with $\delta=1/8$. (a) Pairing correlations.
(b) Spin correlation and single-particle propagator.
(c) The ratio of pair correlations between different directional bonds.
(d) The charge density distribution.
(e) Semi-logarithmic plot of the charge density correlations $N(r)$.
(f) Double-logarithmic plot of the charge density correlations $N(r)$.
A second-order polynomial fitting of $1 / D$ in order to scale $D$ to $D \rightarrow \infty$ has been used to conjecture the true nature of long-distance correlations for $D_{yy}(r)$ (a) and $N(r)$ (e,f).
}
\label{FigS_SC}
\end{figure}

In the main text, we demonstrated the behaviors of different correlation functions when $V_1=-1$ within the $d$-wave SC phase.
In this section, we provide analogous data for a smaller NN attraction at $V_1=-0.5$ and NN repulsion at $V_1=1$, as well as more analysis for the case of $V_1=-1$. Furthermore, we present the functions $D_{yy}(r)$ with respect to $V_1$ across an expanded set of fixed $r$ values (Fig.\ref{FigS_PhaseDiagramSC}), thereby further reinforcing the solidity of a diminishing strength in the pair correlations as $V_1$ increases within the SC phase.
In quasi-one-dimensional cylinders, we explore the presence of SC quasi-long-range order, which is characterized by $D_{\alpha\beta}({r})\sim r^{-\eta_{\mathrm{sc}}}$. Specifically, $\eta_{\mathrm{sc}} < 2$ indicates a divergent superconducting susceptibility in two dimensions as the temperature $T\to 0$.

We show results for a representative value of a smaller NN attraction at $V_1=-0.5$ in Fig.~\ref{FigS_SC}. We present the pair correlations exhibiting a slow power-law decayed rate $\eta_\mathrm{sc}\approx0.65$ (Fig.~\ref{FigS_SC}(a)). Moreover, we find exponentially decaying spin correlations $S(r)$ and single-particle propagator $C(r)$, as illustrated in Fig.~\ref{FigS_SC}(b).
To investigate the SC pairing symmetry, we calculate the pair correlations between different types of bonds. We find that $D_{yx}$ is always negative, while $D_{yy}$ and $D_{xx}$ are always positive. As depicted in Fig.~\ref{FigS_SC}(c), the ratio $D_{y x}/D_{y y}$ is close to $-1$ while $D_{xx}/D_{y y}\approx1$, suggesting the equal amplitude and the opposite signs for the pair correlations between $x$ and $y$ bonds. 
We further examine the charge density distribution $\langle \hat{n}(x,y)\rangle$.
As shown in Fig.~\ref{FigS_SC}(d), we find a uniform charge density, 
consistent with the exponential decay in the charge density correlations $N(r)$ (see Fig.~\ref{FigS_SC}(e) for a semi-logarithmic plot and Fig.~\ref{FigS_SC}(f) for a double-logarithmic plot for comparison).
This behavior in $N(r)$ is similar to that of a larger inter-site attraction such as $V_1=-1$ in the main text (Fig.~\ref{FigS_NrPowerExpVm1}), but different from the case where inter-site electron interaction is absent. 
The inter-site electron repulsion case $V_1>-1/4$ is exemplified by $V_1=1$ in Fig.~\ref{FigS_SCV1}, with $D_{yy}$ and $N(r)$ decaying with a power law in a faster rate $\eta_\mathrm{sc}\approx 1.01$ and in a slower rate $\eta_\mathrm{cdw}\approx 2.48$, respectively, compared with the case of $V_1=-1$ in the main text.
The robustness of these features is confirmed by different values of $L_x$.
Overall, these observed features, along with those for the larger electron attraction $V_1=-1$ in the main text, reinforce the robustness of the quasi-long-range SC order throughout the entire $d$-wave SC phase.

\begin{figure}[H]
\begin{center}
\includegraphics[width=0.58 \textwidth]{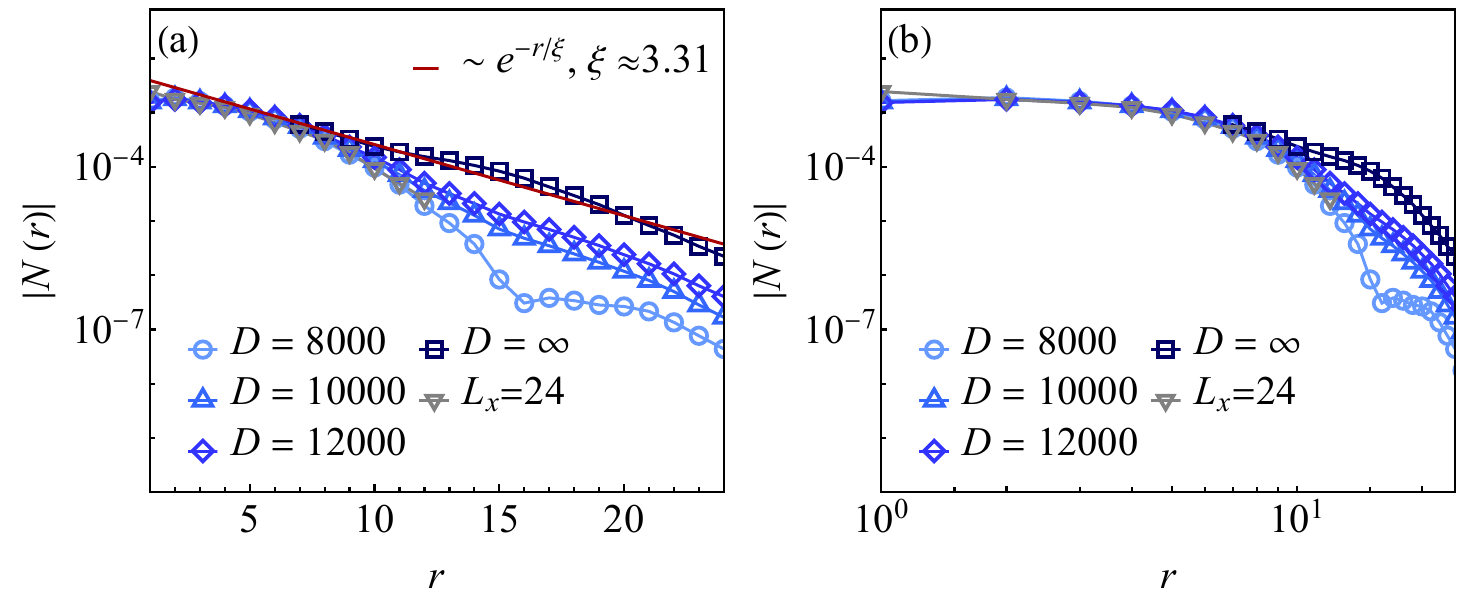}
\end{center}
\par
\renewcommand{\figurename}{Fig.}
\caption{
Semi- (a) and double-logarithmic (b) plot of the charge density correlations $N(r)$ at NN attraction $V_1=-1$ for $N=24\times6$ and $N=48\times6$ with $\delta=1/8$. It appears to be difficult to fit the curves in the double-logarithmic (b) plot using a power-law function. A second-order polynomial fitting of $1 / D$ in order to scale $D$ to $D \rightarrow \infty$ for each distance $r$ has been used to conjecture the true nature of long-distance correlations for both plots. The plot (a) is the same as that in the main text and is included for comparison.
}
\label{FigS_NrPowerExpVm1}
\end{figure}

\begin{figure}[H]
\begin{center}
\includegraphics[width=0.6\textwidth]{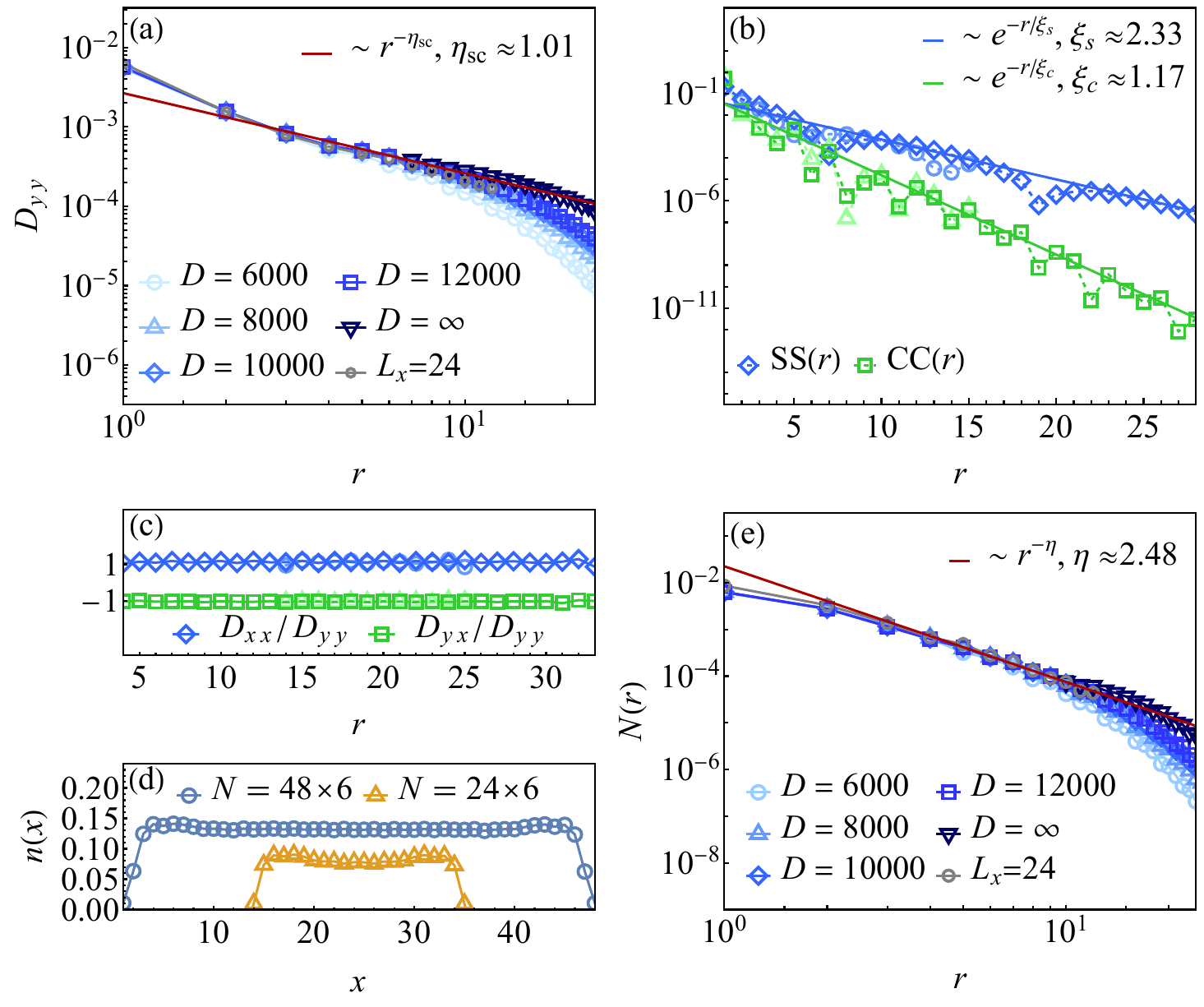}
\end{center}
\par
\renewcommand{\figurename}{Fig.}
\caption{
SC properties at typical NN repulsion $V_1=1$ of the phase for $N=24\times6$ and $N=48\times6$ with $\delta=1/8$.
(a) Double-logarithmic plot of pair correlations $D_{yy}$.
(b) Renormalized spin correlation and single-particle propagator. Darker colors imply $L_x=48$.
(c) The ratio of pair correlations between different directional bonds. Darker colors imply $L_x=48$. Data of $L_x=24$ are offset horizontally for clarity.
(d) The charge density distribution $n(x)$. Data of $L_x=24$ are offset horizontally and vertically for clarity.
(e) Double-logarithmic plot of the charge density correlations $N(r)$.
A second-order polynomial fitting of $1 / D$ in order to scale $D$ to $D \rightarrow \infty$ for each distance $r$ has been used.
}
\label{FigS_SCV1}
\end{figure}

\section{Phase separation for large inter-site attraction}

\begin{figure}[tbp]
\begin{center}
\includegraphics[width=0.6\textwidth]{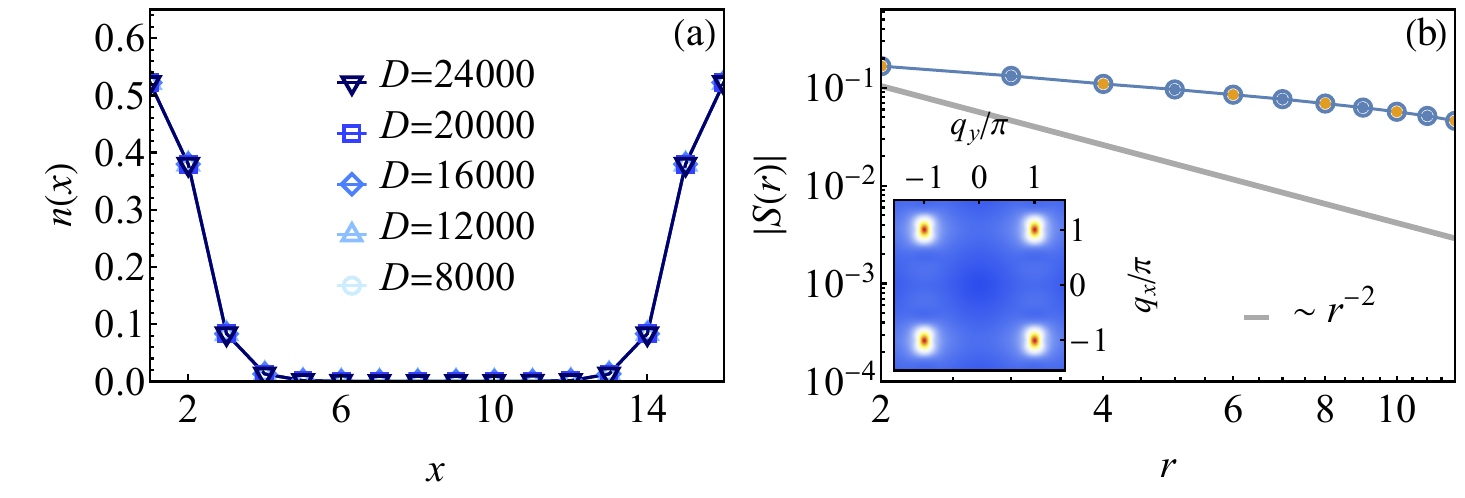}
\end{center}
\par
\renewcommand{\figurename}{Fig.}
\caption{Properties of the phase separation (PS).
(a) The charge density distribution $n(x)$.
(b) Power-law decayed spin correlations $|S(r)|$ with the property $S(r)S(r+1)<0$. Orange dots for $S(r)>0$ and blue dots for $S(r)<0$. Inset, static spin structure factor $S(\mathbf{q})$.
Here, we consider $\delta=1/8$ and the typical parameter $V_1=-3$ of this phase.
}
\label{Fig_PS}
\end{figure}

Our further calculations reveal that larger attractive interaction $V_1\lesssim -1.5$ suppresses SC and drives the system to a phase separation~\cite{Emery1990,Kivelson2003,Gooding1994,Martins2001, Macridin2006,Bejas2014}, 
which refers to the coexistence of insulating magnetic region in the bulk and charge accumulation zone near the boundary [see Fig.~\ref{Fig_PS}(a)]. This phenomenon is demonstrated by examining $\bar{n}_\mathrm{b}$ [see Fig.~1(b) in the main text] defined by
\begin{equation}\label{eq:PS}
  \bar{n}_\mathrm{b} \equiv
  \frac{\sum_{x=1}^{L_x/4}n(x)+\sum_{x=3L_x/4}^{L_x} n(x) -\sum_{x=L_x/4+1}^{3L_x/4-1}n(x)}{\sum_{x=1}^{L_x}n(x)}.
\end{equation}
Here, a value of $\bar{n}_\mathrm{b}\approx1$ serves as a strong indicator of phase separation, while a distinct value suggests the suppression of phase separation. Due to the concentration of charges at the boundary, the bulk spin correlations $S(r)$ resemble the undoped N\`{e}el antiferromagnetic order (AF), i.e., $S(r)$ exhibits a power-law decay
and the AF property $S(r)S(r+1)<0$ [see Fig.~\ref{Fig_PS}(b)].
We also calculate the static spin structure factor $S(\mathbf{q})$, which is defined by the Fourier transformation of spin correlations.
As shown in the inset of Fig.~\ref{Fig_PS}(b), the maximum of $S(\mathbf{q})$ emerges at $\mathbf{q}=(\pi,\pi)$, and the minimum is at $\mathbf{q}=\mathbf{0}$, which is also consistent with the undoped antiferromagnetic phase in the bulk of the system.

\section{Pseudogap-like phase for large inter-site repulsion}

The PG-like phase that emerges in our model for larger inter-site repulsion is characterized by the coexistence of strong CDW, SC, and SDW fluctuations. The intricate interplay of these fluctuations for $V_1=8$ and different system sizes is evident in Fig.~\ref{FigS_PG}(a), where the renormalized $\mathrm{NN}(r)\equiv |N(r)|/\delta^2$ exhibits a prominent amplitude. This amplitude signifies the slightly dominant influence of CDW fluctuations within this phase.
Notably, in Fig.~\ref{FigS_PG}(b), the single particle propagator $|C(r)|$ decays exponentially, {suggesting the charge insulation along $\boldsymbol{e}_x$}.
For the pseudogap-like phase, we have conducted a thorough review of typical numerical results [see Fig.~\ref{FigS_PGV8}], extending calculations to the bond dimension $D=18000$.

\begin{figure}[H]
\begin{center}
\includegraphics[width=0.5\textwidth]{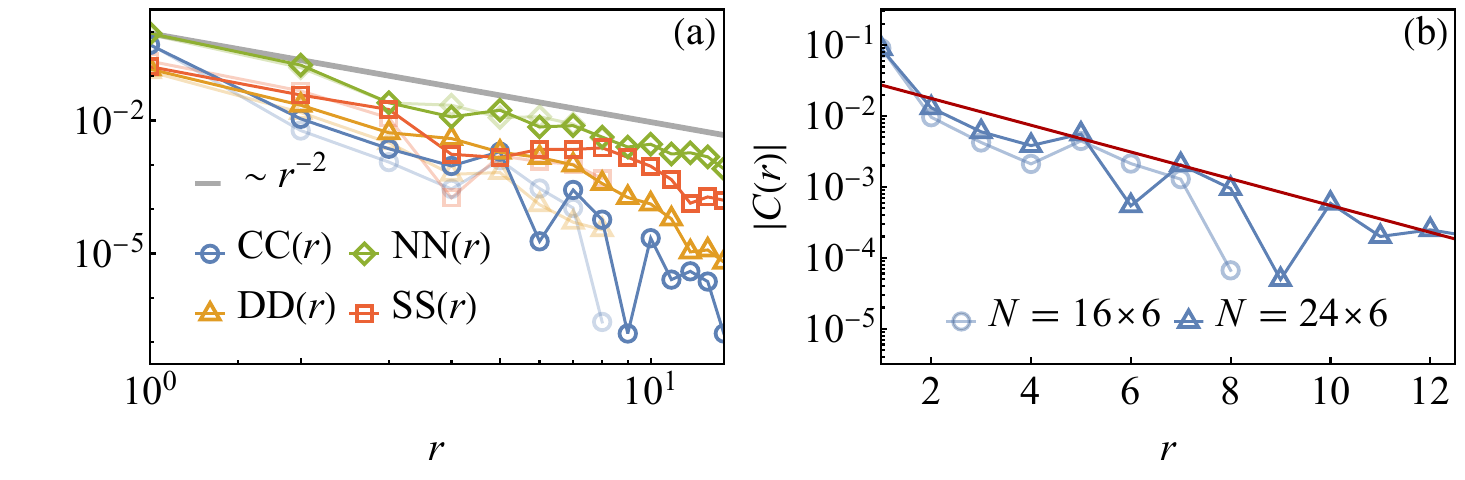}
\end{center}
\par
\renewcommand{\figurename}{Fig.}
\caption{
PG-like properties at a typical parameter $V_1=8$ of the phase for $N=16\times6$ and $N=24\times6$ with $\delta=1/8$.
(a) Various correlations. Renormalization of the correlations is employed to facilitate a direct comparison. 
(b) Exponentially decayed single particle propagator $|C(r)|$. 
Here, darker colors imply $L_x=24$, and we set the reference position $\mathbf{r}_0=(L_x/4,y_0)$ for the correlations.
}
\label{FigS_PG}
\end{figure}

\begin{figure}[H]
\begin{center}
\includegraphics[width=0.5\textwidth]{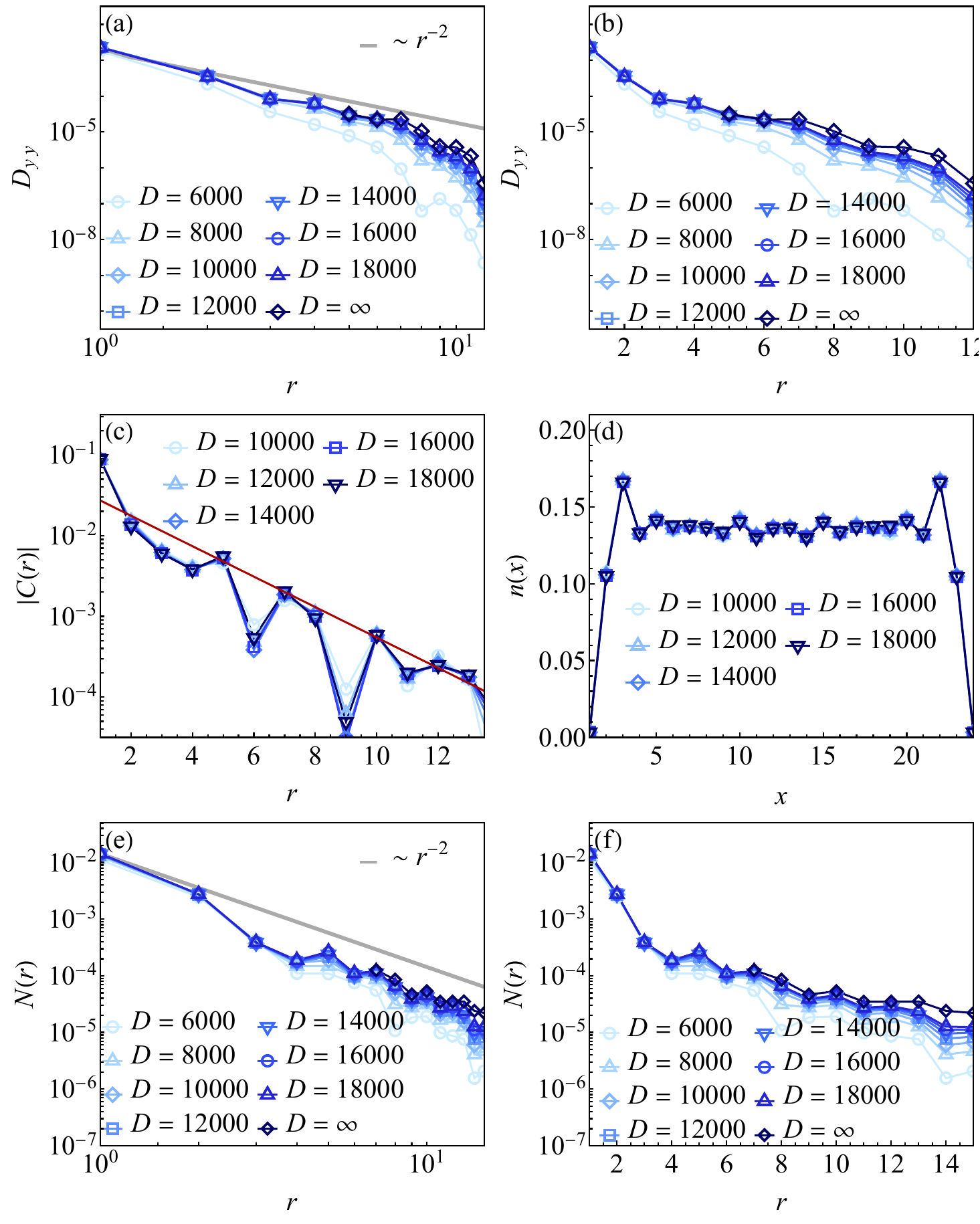}
\end{center}
\par
\renewcommand{\figurename}{Fig.}
\caption{
PG-like properties with respect to different bond dimensions at a typical parameter $V_1=8$ of the phase for $N=24\times6$ with $\delta=1/8$.
(a,b) Double-  (a) and semi-logarithmic (b) plot of the pairing correlations.
(c) Single-particle propagator.
(d) The charge density distribution.
(e,f) Double-  (e) and semi-logarithmic (f) plot of the charge density correlations $N(r)$.
A second-order polynomial fitting of $1 / D$ in order to scale $D$ to $D \rightarrow \infty$ has been used to conjecture the true nature of long-distance correlations for $D_{yy}(r)$ (a,b) and $N(r)$ (e,f).
}
\label{FigS_PGV8}
\end{figure}

\section{Slave-Boson mean field analysis}

In this section, we provide the motivation and details of the slave-boson mean-field approach. Based on our DMRG results, the pair correlations in the SC phase are suppressed by the $V_{ij}$ interaction in the $t$-$J$-$V$ model. In most slave-boson mean-field studies, the interaction term $n_{i}n_{j}$ is usually treated by replacing $n_{i}$ with $1-\delta$. 
To understand the numerical findings, we use the slave-boson mean-field approach to examine the impact of the inter-site interactions $V$ in the $t$-$J$-$V$ model.
The DMRG results presented in the main text reveal a uniform charge density distribution $n(x)$ in the SC phase. This supports the choice of a unit cell in the following mean field analysis, confirming its validity.

In the slave-boson representation, the electron creation operator $c_{i\sigma}^\dagger$ is decomposed into the fermionic spinon $f_{i}^\dagger$ and bosonic holon $b_{i}$,  $c_{i\sigma}^{\dagger}=f_{i\sigma}^{\dagger}b_i$. The Hamiltonian is given by
\begin{equation}
\begin{split}
    H=-t\sum_{\left< ij \right> ,\sigma}{\left( f_{i\sigma}^{\dagger}b_ib_{j}^{\dagger}f_{j\sigma}+f_{j\sigma}^{\dagger}b_jb_{i}^{\dagger}f_{i\sigma} \right)}+J\sum_{\left< ij \right>}{\hat{S}_i\cdot \hat{S}_j}+\left( V-\frac{1}{4}J \right) \sum_{\left< ij \right>}{n_in_j}
\end{split}
\end{equation}
where 
$
\hat{S}_i=\frac{1}{2}\sum_{\alpha ,\beta}{c_{i\alpha}^{\dagger}\hat{\sigma}_{\alpha \beta}c_{j\beta}}$ is the spin operator. $\left< ij \right>$ represents the nearest-neighbor. The single occupation constraint $\sum_{\sigma}{c_{i\sigma}^{\dagger}c_{i\sigma}}\le 1$ 
is written as $\sum_{\sigma}{f_{i\sigma}^{\dagger}f_{i\sigma}} +b_{i}^{\dagger}b_{i}=1$. 

\begin{figure}[tph]
\begin{center}
\includegraphics[width=0.65\textwidth]{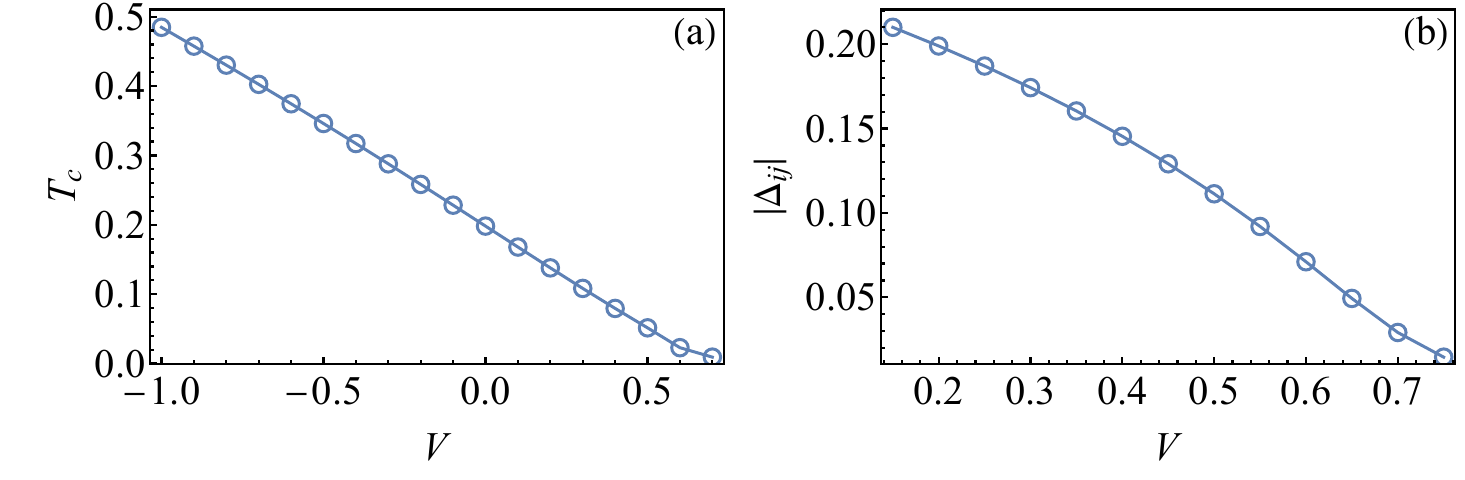}
\end{center}
\par
\renewcommand{\figurename}{Fig.}
\caption{(a) The critical temperature $T_{c}$ dependence of $V$. $T_{c}$ is in units of $J$ for $J/t=1/3$. When shifting the $V$ interaction from attraction to repulsion, a notable decrease of $T_{c}$ is observed. (b) The amplitude of the pairing order parameter $|\Delta_{ij}|$ across a set of $V$ values. Here, the doping level is $\delta=1/8$, and the system size is $N=20\times20$.
}
\label{FigS_MFV}
\end{figure}

To calculate the impact of the $Vn_{i}n_{j}$ term, we define the order parameters:
\begin{equation}
\begin{split}
B&=\left< b_ib_{j}^{\dagger} \right> 
\\
\chi&=\sum_{\sigma}{\left< f_{i\sigma}^{\dagger}f_{j\sigma} \right>}=2\left< f_{i\sigma}^{\dagger}f_{j\sigma} \right> 
\\
\chi^{\ast}&=\sum_{\sigma}{\left< f_{j\sigma}^{\dagger}f_{i\sigma} \right>}=2\left< f_{j\sigma}^{\dagger}f_{i\sigma} \right> 
\\
\Delta&=2\left< f_{j\uparrow}f_{i\downarrow} \right> =-2\left< f_{j\downarrow}f_{i\uparrow} \right> 
\\
\Delta^{\ast}&=2\left< f_{i\downarrow}^{\dagger}f_{j\uparrow}^{\dagger} \right> =-\,2\left< f_{i\uparrow}^{\dagger}f_{j\downarrow}^{\dagger} \right> 
\end{split}
\end{equation}
The mean-field Hamiltonian can be decoupled in this form: 
\begin{equation}
\begin{split}
H^{MF}=H_{t}^{MF}+H_{JV}^{MF}-\mu \sum_{i,\sigma}{f_{i\sigma}^{\dagger}f_{i\sigma}}+\lambda \sum_i{\left( \sum_{\sigma}{f_{i\sigma}^{\dagger}f_{i\sigma}}+b_{i}^{\dagger}b_i-1 \right)}
\end{split}
\end{equation}
where
\begin{equation}
\begin{split}
H_{t}^{MF}&=-tB\sum_{\left< ij \right>}{\left( \sum_{\sigma}{f_{i\sigma}^{\dagger}f_{j\sigma}}+\sum_{\sigma}{f_{j\sigma}^{\dagger}f_{i\sigma}} \right)}-t\chi \sum_{\left< ij \right>}{\left( b_{j}^{\dagger}b_i+b_{i}^{\dagger}b_j \right)}+2t\sum_{\left< ij \right>}{\chi B}\\
\end{split}
\end{equation}
\begin{equation}\label{eqS:MFH}
\begin{split}
H_{JV}^{MF}=&-\left( \frac{1}{4}J+\frac{1}{2}V \right) \chi \sum_{\left< ij \right>}{\left( \sum_{\sigma}{f_{i\sigma}^{\dagger}f_{j\sigma}}+\sum_{\sigma}{f_{j\sigma}^{\dagger}f_{i\sigma}}-\chi ^2 \right)}
\\
&-{\left( \frac{1}{2}J-\frac{1}{2}V \right)} \sum_{\left< ij \right>}{\left[ \Delta ^{\ast}\left( f_{j\uparrow}f_{i\downarrow}-f_{j\downarrow}f_{i\uparrow} \right) +\Delta \left( f_{i\downarrow}^{\dagger}f_{j\uparrow}^{\dagger}-f_{i\uparrow}^{\dagger}f_{j\downarrow}^{\dagger} \right) -\Delta ^{\ast}\Delta \right]}
\end{split}
\end{equation}
$\mu$ is the chemical potential, $\lambda$ is the Lagrange multiplier.
After the Fourier transformation and Bogoliubov transformation, we obtain the free energy:
\begin{equation}
\begin{split}
    F=&-\frac{2}{\beta}\sum_k{\ln  \cosh \left( \frac{\beta E_k}{2} \right)}+\sum_k{\varepsilon _k}+\frac{1}{\beta}\sum_k{\ln \left( 1-e^{-\beta \omega _k} \right)}\\
    &+4NtB\chi+N\left( \frac{1}{2}J+V \right) \chi^2+N\left( \frac{1}{2}J-\frac{1}{2}V \right) \left( \Delta _{x}^{2}+\Delta _{y}^{2}\right)-N\lambda 
\end{split}
\end{equation}
where
\begin{equation}
\begin{split}
    \varepsilon _k=&-\left( tB+\frac{1}{4}J\chi +\frac{1}{2}V\chi \right) K\left( k \right) -\mu +\lambda\\
    K\left( k \right) =&2\left( \cos \left( k_x \right) +\cos \left( k_y \right) \right)\\
    \omega _k=&-t\chi K\left( k \right) +\lambda\\
    E_k=&\sqrt{\varepsilon _{k}^{2}+\Delta _{k}^{2}}\\
    \Delta _k=&-\left( J-V \right) \left( \Delta _x\cos \left( k_x \right) +\Delta _y\cos \left( k_y \right) \right) 
\end{split}
\end{equation}
Finally, we minimize the free energy and obtain a set of self-consistent equations:
\begin{equation}
\begin{split}
    \delta&=\frac{1}{N}\sum_k{\frac{1}{e^{\beta \omega _k}-1}}\\
    1-\delta&=\frac{1}{N}\sum_k{\left[ 1-\frac{\varepsilon _k}{E_k}\tanh \left( \frac{\beta E_k}{2} \right) \right]}\\
    B&=\frac{1}{4N}\sum_k{\frac{K\left( k \right)}{e^{\beta \omega _k}-1}}\\
    \chi&=\frac{1}{4N}\sum_k{K\left( k \right) \left[ 1-\frac{\varepsilon _k}{E_k}\tanh \left( \frac{\beta E_k}{2} \right) \right]}\\
    \Delta_\alpha&=\frac{\left(J-V\right)}{N}\sum_k{\frac{\tanh \left( \frac{\beta E_k}{2} \right)}{E_k}\left( \Delta_x\cos k_x+\Delta_y\cos k_y \right)\cos k_\alpha } , \quad\alpha = x,y.
\end{split}
\end{equation}

As discussed above, the $V n_{i}n_{j}$ term represents the Coulomb repulsive or attractive interaction, depending on the sign of $V$.
To understand the numerical findings of the role of such a term in superconductivity, we choose the doping level $\delta=1/8$, and calculate the critical temperature $T_{c}$ and the amplitude of the pairing order parameter $|\Delta_{ij}|$ across a set of $V$ values. The critical temperature $T_{c}$ is determined by solving the self-consistent equations. Here we choose $N=10\times 10\times 10$ lattice sites. We examine the critical temperatures of RVB ($T_\mathrm{RVB}$) and Bose condensation of holons ($T_\mathrm{BE}$) with interlayer electron transfer $t_{z}=0.1t$. In the doping level $\delta=1/8$, we find $T_c=T_\mathrm{RVB}<T_\mathrm{BE}$. Next, we calculate the $T_c$ for $N=20\times20$. As illustrated in Fig.~\ref{FigS_MFV}(a), $T_c$ decreases with the increase of $V$  from attraction to repulsion, consistent with the DMRG observations. Moreover, $|\Delta_{ij}|$ also decreases with increasing $V$ [see Fig.~\ref{FigS_MFV}(b)]. These mean-field results could be clearly observed from the opposite sign of $J$ and $V$ before the cooper pair order parameters in Eq.~\eqref{eqS:MFH}, where the positive $V$ is unfavorable for pairing, unlike the attractive $V$.

\section{Néel antiferromagnetic order and incommensurate spin-density-wave phase}

In this section, we supplement more data of static spin structure factor $S(\mathbf{q})$ and spin correlations $S(r)$ in Eq.~\eqref{eqS:Sr} regarding N\`{e}el antiferromagnetic order (AF) at light doping and incommensurate spin density wave (iSDW) for large doping with various inter-site interaction $V_1$. The static spin structure factor $S(\mathbf{q})$ is defined by 
\begin{equation}
   S(\mathbf{q})\equiv\sum_{\mathbf{i},\mathbf{j}} 
\langle {\mathbf{S}_\mathbf{i}\cdot \mathbf {S}_\mathbf{j}} \rangle 
e^{i \mathbf{q}\cdot (\mathbf{i}-\mathbf{j})}/N ,
\end{equation}
with its peak  at certain wave vectors $\mathbf{q}=\mathbf{q}_0$ signaling magnetic order.  For the AF on a square lattice, the peaks are located at $\mathbf{q}_0=(\pm \pi,\pm \pi)$.

\begin{figure}[H]
\begin{center}
\includegraphics[width=0.7\textwidth]{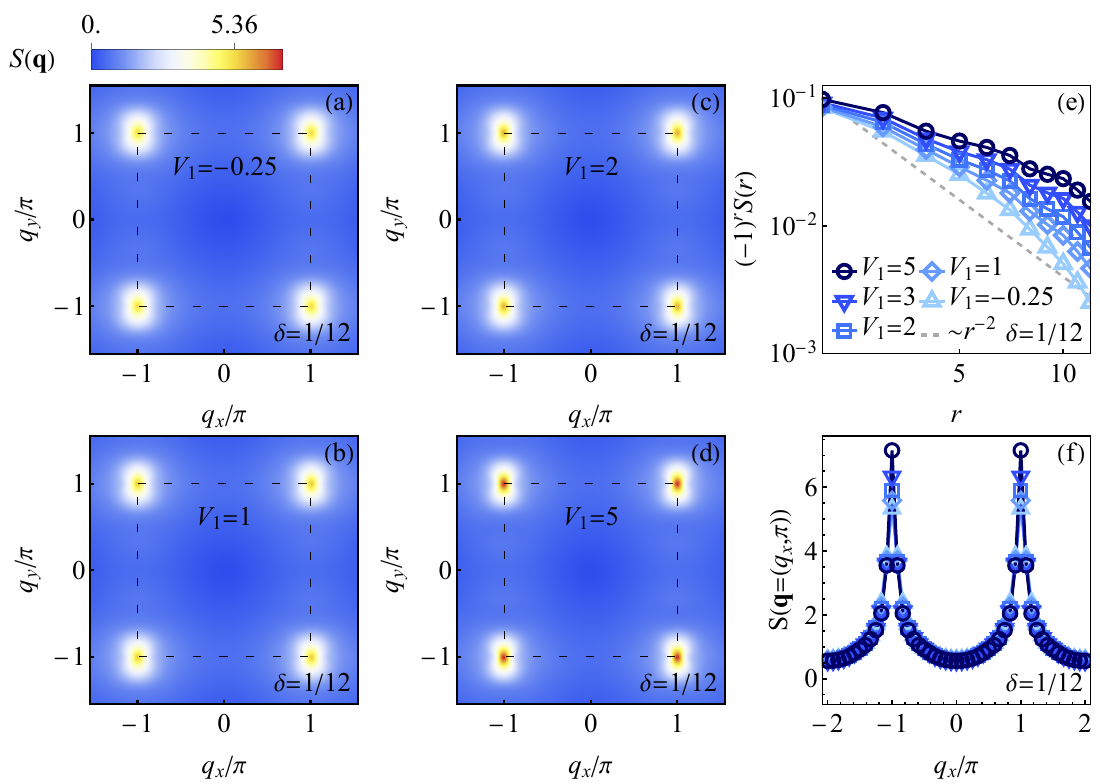}
\end{center}
\par
\renewcommand{\figurename}{Fig.}
\caption{
Evolution of the magnetic structure across a range of inter-site interactions $V_1$ with $N=24\times 6$ at $\delta=1/12$.
(a-d)  The contour plot of the static spin structure factor $S(\mathbf{q})$ with various $V_1$.
Squares with black dashed edges in panels (a-d) denote the Brillouin zone. Interpolation has been applied in the contour plot. 
(e) Amplitude of spin correlations with various $V_1$. Both panels (e) and (f) are labeled identically. 
(f) Line-cut plot of $S(\mathbf{q})$ along the momentum path $\mathbf{q}=(q_x,\pi)$. 
The energy unit is chosen as $J_1$.
}
\label{FigS_SDW}
\end{figure}

\begin{figure}[H]
\begin{center}
\includegraphics[width=0.7\textwidth]{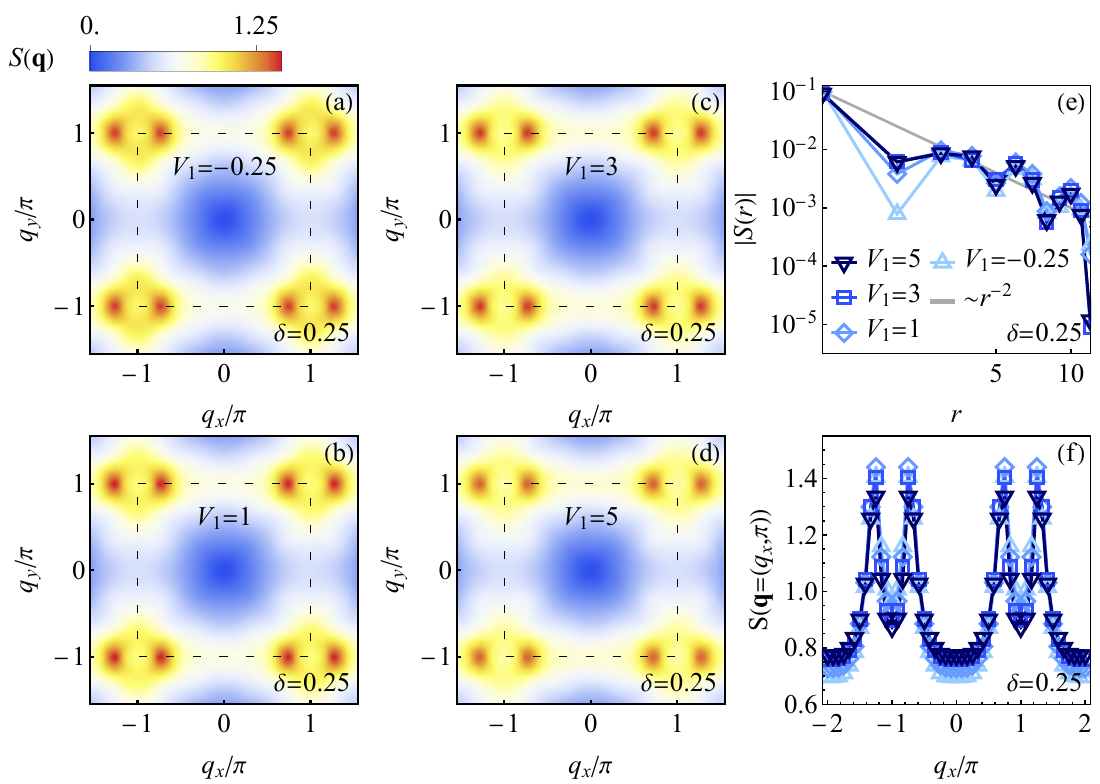}
\end{center}
\par
\renewcommand{\figurename}{Fig.}
\caption{
Evolution of the magnetic structure across a range of inter-site interactions $V_1$ with $N=24\times 6$ at $\delta=1/4$.
(a-d)  The contour plot of the static spin structure factor $S(\mathbf{q})$ with various $V_1$.
Squares with black dashed edges in panels (a-d) denote the Brillouin zone. Interpolation has been applied in the contour plot. 
(e) Amplitude of spin correlations $|S(r)|$ with various $V_1$. Both panels (e) and (f) are labeled identically. 
(f) Line-cut plot of $S(\mathbf{q})$ along the momentum path $\mathbf{q}=(q_x,\pi)$. 
The energy unit is chosen as $J_1$.
}
\label{FigS_iSDW}
\end{figure}

We demonstrate the impact of inter-site interactions on AF at a representative light doping $\delta=1/12$  in Fig.~\ref{FigS_SDW}.
In the limit of standard $t$-$J$ model, i.e., $V_1=-0.25$,  $|S(r)|$ decay at a rate faster than $\sim r^{-2}$. However, with the increase of inter-site interactions,  $|S(r)|$  exhibits a slower decay, approaching power-law behavior with stronger repulsion. Meanwhile, the peaks of $S(\mathbf{q})$ stabilize at $\mathbf{q}_0=(\pm\pi,\pm\pi)$ [see Figs.~\ref{FigS_SDW}(a-d,f)] and intensify [see Figs.~\ref{FigS_SDW}(f)], thereby demonstrating a preference for AF by $V_1$ at light doping. 
Such preference for AF by inter-site repulsion is intriguing for future analytical study.

We identify the ground state at $\delta=1/4$ as iSDW, since the peaks of $S(\mathbf{q})$ slightly split around the momentum $(\pm\pi,\pm\pi)$, as demonstrated in Figs.~\ref{FigS_iSDW}(a-d,f), and all the decay rates shown in Fig.~\ref{FigS_iSDW}(e) are slightly slower than  $\sim r^{-2}$, indicating the presence of a quasi-long-range SDW order. 
The impact of inter-site interactions on iSDW at higher doping levels $\delta=1/4$ is less pronounced, as shown in Fig.\ref{FigS_iSDW}(e). 
Notably, the amplitudes of $|S(r=L_x/2)|$ at $\delta=1/4$ [see Fig.~\ref{FigS_iSDW}(e)] are significantly smaller than those at $\delta=1/12$ [see Fig.~\ref{FigS_SDW}(e)], thereby masking the signature of iSDW in the Fig.~2(b) of the main text.

\end{document}